\documentclass[prb,showpacs,amssymb,superscriptaddress]{revtex4}
\usepackage{txfonts}
\usepackage{amssymb}
\usepackage{graphicx}
\usepackage{subfigure}
\usepackage{dcolumn}
\usepackage{bm}

\begin{document}
\title{Electronic structure and Fermi surface character of LaONiP from first principles}
\author{Wei-Bing Zhang}
\affiliation{School of Chemistry and Chemical Engineering, Guangxi University, Nanning, 530004, China.}%
\affiliation{Department of Physics and Key Laboratory of Low
Dimensional Materials $\&$ Application Technology of Ministry of
Education, Xiangtan University, Hunan Province, 411105, China.}
\author{Xiao-Bing Xiao}
\affiliation{Department of Physics and Key Laboratory of Low
Dimensional Materials $\&$ Application Technology of Ministry of
Education, Xiangtan University, Hunan Province, 411105, China.}
\author{Wei-Yang Yu}
\affiliation{Department of Physics and Key Laboratory of Low
Dimensional Materials $\&$ Application Technology of Ministry of
Education, Xiangtan University, Hunan Province, 411105, China.}
\author{Na Wang}
\affiliation{Department of Physics and Key Laboratory of Low
Dimensional Materials $\&$ Application Technology of Ministry of
Education, Xiangtan University, Hunan Province, 411105, China.}
\author{Bi-Yu Tang}
\email{tangbiyu@xtu.edu.cn}
\affiliation{School of Chemistry and Chemical Engineering, Guangxi University, Nanning, 530004, China.}%
\affiliation{Department of Physics and Key Laboratory of Low
Dimensional Materials $\&$ Application Technology of Ministry of
Education, Xiangtan University, Hunan Province, 411105, China.}

\begin{abstract}
 Based on First-principles calculation, we have investigated
electronic structure of a ZrCuSiAs structured superconductor LaONiP.
The density of states, band structures and Fermi surfaces have been
given in detail. Our results indicate that the bonding of the La-O
and Ni-P is strongly covalent whereas binding property between the
LaO and NiP blocks is mostly ionic. It's also found that four bands
are across the Fermi level and the corresponding Fermi surfaces all
have a two-dimensional character. In addition, we also give the band
decomposed charge density, which suggests that orbital components of
Fermi surfaces are more complicated than cuprate superconductors.
\end{abstract}

 \maketitle

\section{\label{sec:Intro}Introduction}

Quaternary rare earth transition metal phosphide oxides with the
tetragonal ZrCuSiAs structure \cite{ZrCuSiAs} are known for more
than twenty years\cite{JAC-1995}. Numerous quaternary phosphide
oxides have been reported, including the actinoidand
copper-containing compounds UOCuP,\cite{UcuPO} ThOCuP
\cite{ThCuPO}as well as the series of lanthanoid (Ln) compounds
LnOFeP\cite{JAC-1995}, LnORuP\cite{JAC-1995}, LnOCoP\cite{JAC-1995},
LnOOsP, LnOZnP\cite{LaZnOP,electronicLaZnOP} and
LnOMnP\cite{LaMnPO1997qem}.

However, most of interests are limited to the synthesization and
structural analysis, only few physical properties have been
investigated for this compound series. Because of open-shell
structure and electronic correlation of 3d electrons, the ZrCuSiAs
type transition metal phosphide oxides are expected to display
interesting magnetic and electronic phenomena. With the same crystal
structure, the different groundstates including antiferromagnetic
insulator \cite{kabbour2005rdn}, semiconductor and metal
\cite{electronicLaZnOP}have been found in the ZrCuSiAs type
compounds. Moreover, LaOFeP \cite{jacs_LaFeOP,LaOFeP_SSt} has been
found to be a type II superconductor with $T_c$ of 3.2\~{}6.5 K ,
which leads to renewed interest in studying physical properties in
this class of materials. Later, $T_c$ around 26 K has been found in
F-doped \cite{LaOFeAs_F}and Sr-doped \cite{LaOFeAs_Sr} LaOFeAs. Very
recently, up to $T_c$=55 K was reached  in LnO$_{1-x}$F${_x}$FeAs
through replacement of La by Ce\cite{Ce-2008}, Pr\cite{Pr-2008},
Nd\cite{Nd-2008}, Sm\cite{sm-2008,sm1-2008},and Gd
\cite{Gd-2008-51}. Ferromagnetic Kondo lattice systems found in
Ce-based phosphide oxides CeOTP
(T=Ru,Os)\cite{CeRuPO,krellner2007scg} also attract much attentions.

Recently, a new compound LaONiP with ZrCuSiAs structure has been
reported by Watanabe \emph{et al.}\cite{LaNiO_IC}, and it has also
been synthesized by Tegel \emph{et al.}\cite{LaNiPO_SSC} with
different experimental method.  It has been identified as a
superconductor with a critical temperature $T_c$ of 3 and 4.3 K,
respectively. Superconductivity was also found in LaONiAs and
LaO$_{1-x}$F${_x}$NiAs\cite{LaONiAs}. LaONiP is the first nickel
phosphide oxide and the second superconducting compound in
LaOMP(M=transition metal). It seems that the superconductivity may
be a common phenomena in this family. Although this $T_c$ is lower
than that of the copper oxides, because of possible different
underlying mechanism, the discovery of superconductivity in the new
material systems can provide valuable knowledge for understanding of
superconduction and for finding another superconductor. Thus, it is
crucial to clarify the underlying mechanism of novel
superconductivity in this series of compounds. Whereas the
understanding of electronic structure especially the Fermi surface
character is a basic step to obtain insight into the
superconductivity mechanism.

In present calculation, we perform the systematic investigation for
LaONiP based on density functional theory. The detailed structural
parameters, electronic structures and Fermi surfaces were given. The
bonding properties are also discussed in combination with the
concept of two-dimensional (2D) building blocks and a Bader analysis
of the charge density. The band decomposed charge density is also
provided to analyze the orbital character of each Fermi surface.
\section{\label{sec:Compu}Computational Details}
The present calculations have been performed using the Vienna Ab
initio Simulation Package (VASP) code
\cite{vasp_CMS,vaspPhysRevB.54.11169} within projector
augmented-wave (PAW) method, \cite{vasp_PAW,vasppawPhysRevB.59.1758}
general gradient approximations (GGA) \cite{PW91}were used in the
present calculations. The La($5s^25p^65d^16s^2$), Ni($3d^94s^1$),
P($3s^23p^3$), O($2s^22p^4$) are treated as valence states. To
ensure enough convergence, the energy cutoff was chosen to be 600
eV, and the Brillouin zone was sampled with a mesh of $16\times
16\times 8 $ \emph k points generated by the scheme of
Monkhorst-Pack \cite{VASP_MP}. A first order Methfessel-Paxton
method with $\sigma =0.2 eV $ was used for relaxation. The crystal
cell and internal parameters were optimized using the conjugate
gradient method until the total forces on each ion less than 0.01
eV/\AA. Then density of states (DOS) calculations were performed
using the tetrahedron method with the Bl\"ochl
corrections\cite{PhysRevB.49.16223}. The Fermi surfaces and 3D
charge density iso-surfaces have been drawn by
\emph{Xcrysden}\cite{xcrysden}.
\section{\label{sec:results} Results and Discussion}
\subsection{\label{structure}Crystal Structure}
\begin{figure}
\centering
\includegraphics[width=0.42\textwidth]{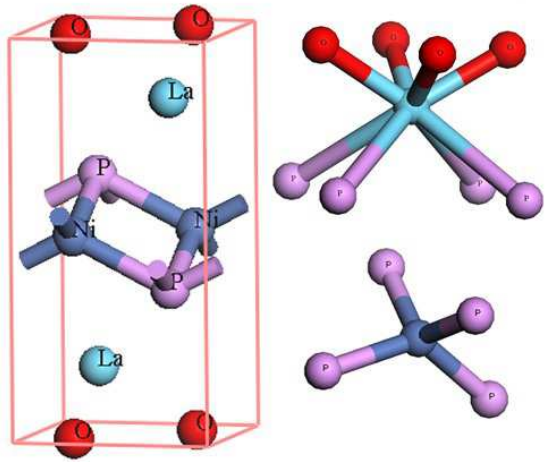}
\caption{\label{fig:strucure}(Color online)Crystal structure of
LaONiP and near-neighbour coordinations of La and Ni atom. The
element symbols have been labeled in the figure.}
\end{figure}

LaONiP  crystallizes in the tetragonal ZrCuSiAs structure (space
group $P4/nmm$) with the Ni atom  at 2a (0.75, 0.25, 0), O at 2b
(0.75, 0.25, 0.5), La at 2c (0.25, 0.25, $z_1$) and P at 2c(0.25,
0.25, $z_2$). It has a layered structure composed of an alternating
stack of LaO and NiP block. The crystal structure is therefore
rather simple with eight atoms (two formula units) in the unit cell
as shown in Fig.~\ref{fig:strucure}. Lanthanum is eightfold
coordinated by four oxygen and four phosphorus atoms. The nickel
atoms are tetrahedrally coordinated by four phosphorus atoms,
forming a distorted tetrahedra with two different P-Ni-P
angles.\cite{LaNiO_IC,LaNiPO_SSC} And they also have four
neighboring nickel atoms within the same layer.

Table~\ref{tab:table1} gives the optimal structural parameter of
LaONiP together with available experimental results. The
experimental results are very consistent with each other. The
present calculation is also in agreement with experiment, which
indicates that the calculation is accurate. The slight difference
between experiment and calculation may be induced by the fact that
the LDA/GGA calculation cannot describe the  electronic correlation
between 3d electrons of Ni atom accurately. It should be noticed
that the electronic structure \cite{ref1,AsFe1}and magnetic
properties \cite{ref2} in case of LaOFeAs are very sensitive to
small change of internal coordinates, as shown in
Ref.~\onlinecite{ref1,AsFe1,ref2}. It seems that in order to give a
more reasonable description, further calculation including the
electronic correlation should be performed.

\begin{table}
\centering
 \caption{\label{tab:table1}Comparison of experimental and calculated crystallographic Parameters of LaONiP.}
\begin{ruledtabular}
\centering
\begin{tabular}{ccccc}
&$a$ (\AA)&$c$ (\AA)&$z_1$&$z_2$\\
 \hline
Expt.1\footnote{Ref.~\onlinecite{LaNiO_IC}.}& 4.0461 & 8.100 & 0.1531 &0.626 \\
Expt.2\footnote{Ref.~\onlinecite{LaNiPO_SSC}.}&4.0453&8.1054&0.1519&0.6257\\
Calculation&4.0353&8.2168&0.1513&0.6247 \\
\end{tabular}
\end{ruledtabular}
\end{table}

\subsection{\label{sec:electronic}Density of States and Binding Properties }
\begin{figure}
\centering
 \subfigure[]{\label{fig2:a}\includegraphics[width=0.530\textwidth]{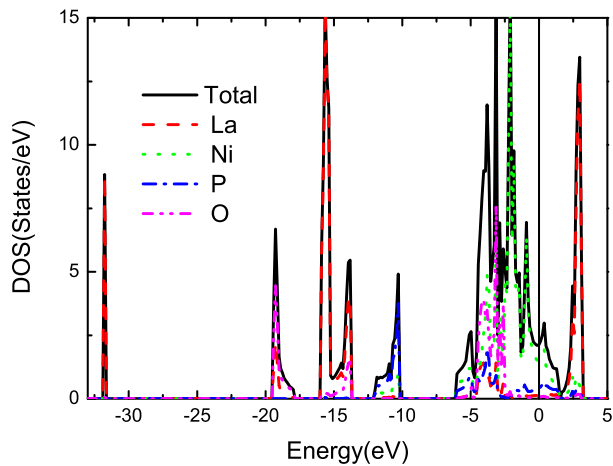}}\\
 \subfigure[]{\label{fig2:b}\includegraphics[width=0.530\textwidth]{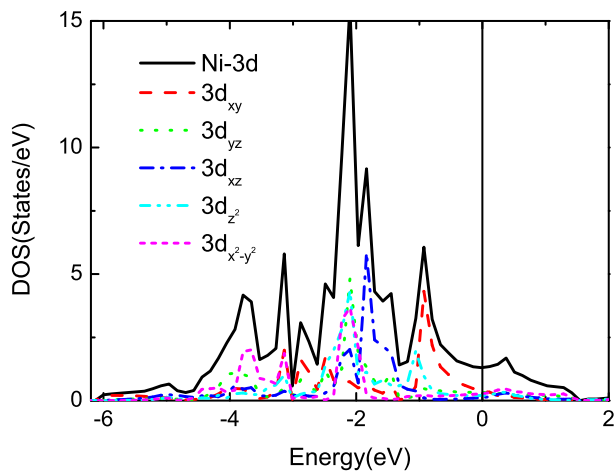}}\\
 \subfigure[]{\label{fig2:c}\includegraphics[width=0.530\textwidth]{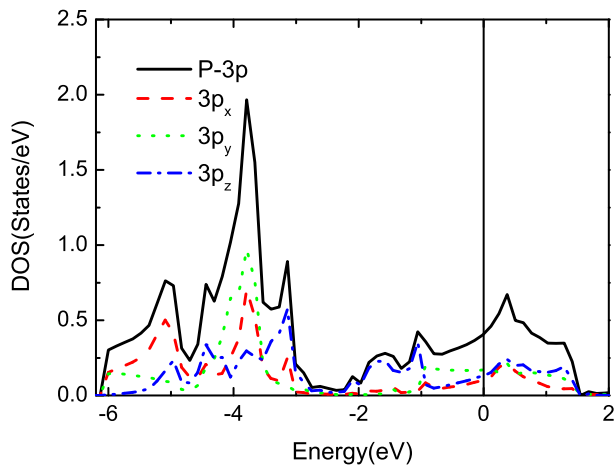}}
\caption{\label{fig2}(Color online) Total density of states and
Partial density of states for each atomic species(a) and the orbit
resolved density of states for Ni 3d states(b) and P 3p states(c) of
LaONiP. The Fermi level is at zero energy.}
\end{figure}

Fig.~\ref{fig2} shows the calculated total and partial DOS of
LaONiP. We can see the states below -13 eV is contributed by La and
O atom. And the La-p and O-s shows strong hybridization with each
other under the range -20 to -13 eV, indicating strong La-O bonding.
The energy range from -12 to -10 eV is predominated by the P and Ni
atoms with only a weak mixture of La atom. Whereas all atoms are
contributing to the DOS at the energy range -6 to -3 eV. It should
be noticed that from -2 to 2 eV, the 3d states of Ni atoms dominate
the DOS, together with a slight contribution from P-3p states, while
the La and O almost may be neglected. These results indicate that
there are a significant overlap between the orbit in La-O and Ni-P
and the bonding differs considerably from ionic ones.

Recently, the electronic structure of
LaOFePn(Pn=P,As)\cite{LaFeOPfermi,AsFe1,corralted,ref1,ref2}have
been investigated extensively. These calculations predict that the
Fermi energy lies just above a peak in the DOS. Thus, LaOFePn have a
very steep and negative slope of the density of states (DOS) at the
Fermi level, which drives the system close to a magnetic
instability. Compared with iron-based compound, we find that the
shape of DOS in case of LaONiP is similar. However, because of
Ni$^{2+}$ (3d$^{8}$) contributes two more electrons than Fe$^{2+}$
(3d$^{6}$)in LaOFePn,  Fermi level is lifted up in case of LaONiP
and the densities of states don't decrease monotonously with energy,
which may put LaONiP away from magnetic instability.

As shown in above DOS analysis, we can find that the states of Ni
and P atom predominate DOS near Fermi level, which suggests that the
superconductivity origins from the contribution of the Ni 3d and P
3p as shown in Fig.~\ref{fig2:a}. The layered copper-based
superconductors have been extensively studied for several decades.
In cuprate superconductors, Cu$^{2+}$ occupies a planar 4-fold
square site and the charge carriers at the Fermi level are driven by
the $d_{x^2-y^2}$ orbit. While Ni$^{2+}$ in LaONiP occupies a
tetrahedral site coordinated with four P$^{3-}$ ions. Such a marked
difference in the coordination structure between the LaONiP and
cuprate superconductor is expected to lead to the different
mechanism of superconductivity. The Ni ion in LaONiP is formally in
a 3$d^8$ configuration. In a tetrahedral crystal field, The Ni d
bands will split into a lower lying e and an upper lying t$_2$
orbit. However, because of the distorted tetrahedral as shown in
Fig.~\ref{fig2:b}, a clear separation in energy of the d orbit in a
lower e ($3d_{z^2}$, $3d_{x^2-y^2}$) and higher t$_2$($3d_{xy}$,
$3d_{yz}$, $3d_{xz}$) set is not seen in the present calculation.
Whereas  $3d_{xy}$ and $3d_{z^2}$ contributes more to DOS near the
Fermi surface, which may play a more important role in
superconductivity. In addition, from Fig.~\ref{fig2:b}and
Fig.~\ref{fig2:c}, all five 3d orbits of Ni atom and three p orbits
of P atom has comparable contribution to DOS at the Fermi level.

\begin{figure*}
\centering
\includegraphics{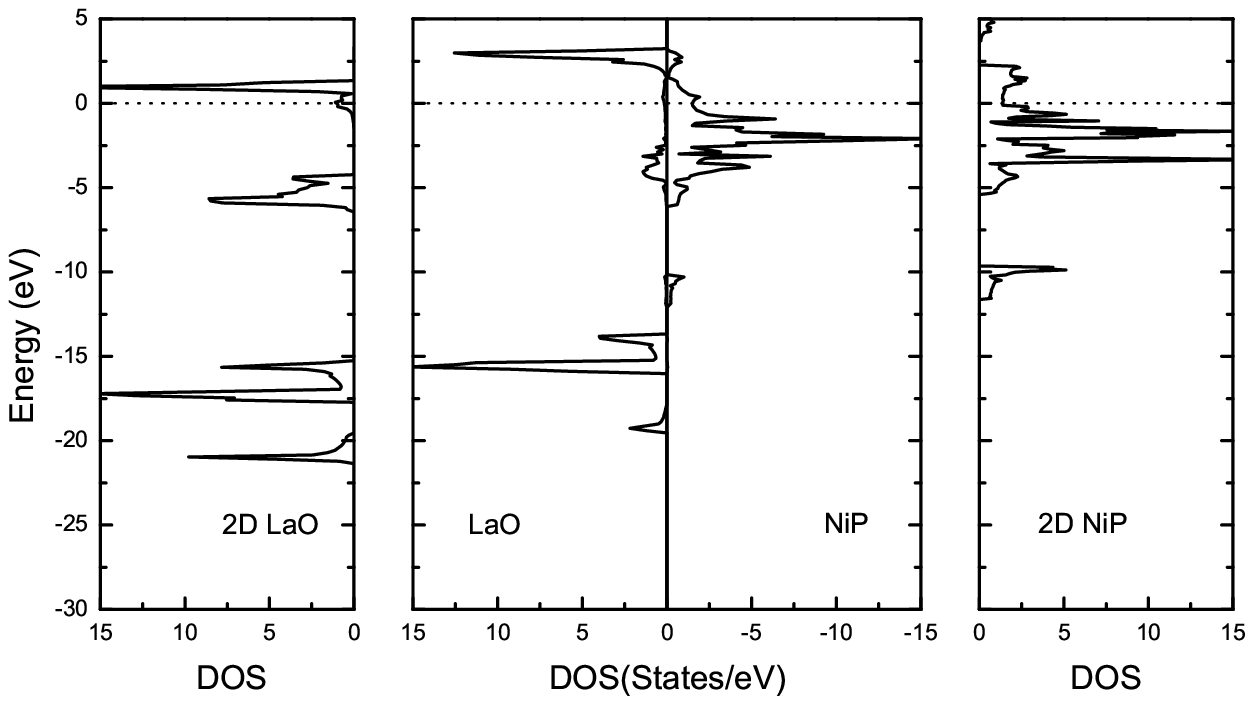}
\caption{\label{fig:epsart}(Color online) Comparison of the DOS of
the LaO(left) and NiP (right) blocks calculated as isolated entities
with the PDOS for LaO (center left) and NiP (center right) in the
LaONiP solid.}
\end{figure*}

We also analysis the DOS using the concept of two-dimensional (2D)
building blocks\cite{10.1038/nature01650,BanglinChen02092001,
CarioL.cm048180p, KabbourH.ic051592v}. It will be helpful to
understand how charge transfer and hybridization effects will affect
their electronic and binding properties when building the solid from
these separated blocks. From the Fig.~\ref{fig:epsart}, we can see
that the LaO states are pushed up in energy by approximately 2 eV ,
and becoming insulating in the solid but the isolated LaO layer is
metallic, which indicate there are electron transfers from the LaO
to NiP when building the solid. From the figure, we also conclude
that the LaO-FeP interaction is strongly ionic character with weak
hybridization under energy range of -5 and -2 eV.

As shown above, the qualitative electronic structure and binding
properties is very similar to the results of LaOFeP give in
Ref.~\onlinecite{LaFeOPfermi}.  In order to understand the binding
properties more quantificationally, we also performed a Bader
analysis \cite{bader1990amq,bader2006far} of the charge density. The
core charge missed in general pseudopotentials  method has also been
added to perform this analysis. Table~\ref{tab:charges} gives the
charges of each atom specie using bader analysis together with the
pure ionic picture. We can notice that the La-O and Ni-P bonds
differ considerably from ionic character, whereas the bonding
between the LaO and FeP blocks is mostly ionic. It is consistent
with the previous DOS analysis. Compared with the result of LaOFeP
\cite{LaFeOPfermi}, we also find that the deviation of ionic
character between Ni and P is larger than the case of LaOFeP, which
suggests the covalency between 3d metal and P atoms is stronger in
LaONiP. The weak p-d hybridization between Fe and As is also found
in case of LaOFeAs\cite{AsFe1,corralted,ref1}.

\begin{table*}
 \caption{\label{tab:charges} Electronic charges belonging
to each atomic specie obtained by Bader analysis, compared with a
purely ionic picture. The results of LaOFeP taken from the
Ref.~\onlinecite{LaFeOPfermi} are included for comparison. }
\begin{ruledtabular}
\begin{tabular}{ccccccc}
&La&M&P&O&LaO&MP\\
 \hline
Bader(LaONiP) & 9.0902 & 9.9435 &5.6841& 7.2822 &16.3724&15.6276 \\
ionic picture(LaONiP) &8(La$^{3+}$)&8(Ni$^{2+}$)&8(P$^{3-}$)&8(O$^{2-}$)&16(LaO$^+$)&16(NiP$^-$)\\
LaOFeP\footnote{Ref.~\onlinecite{LaFeOPfermi}}&9.05&7.59&6.05&7.31&16.36&13.64\\
\end{tabular}
\end{ruledtabular}
\end{table*}

\subsection{\label{sec:Fermi}Band Structure and Fermi Surface }
\begin{figure}
\centering
\includegraphics[width=0.45\textwidth]{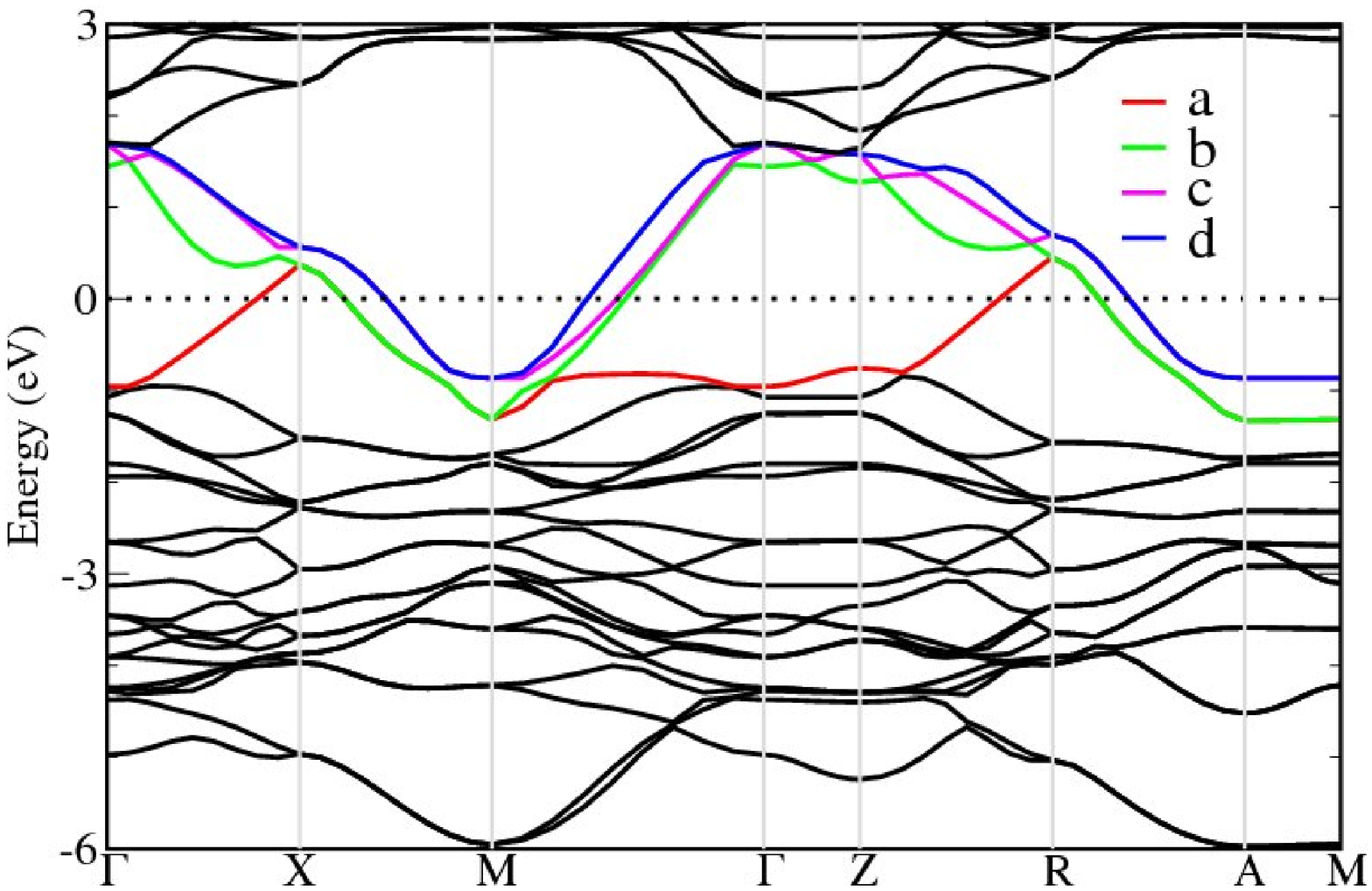}
\caption{\label{fig:band}(Color online) The calculated band
structure of LaONiP. The Fermi level is at zero energy and marked by
a horizontal dashed line. Four bands across the Fermi level are
indicated in figure and marked with different colors.}
\end{figure}

\begin{figure*}
\centering
 \subfigure[]{\label{fig:subfig:a}\includegraphics[width=0.450\textwidth]{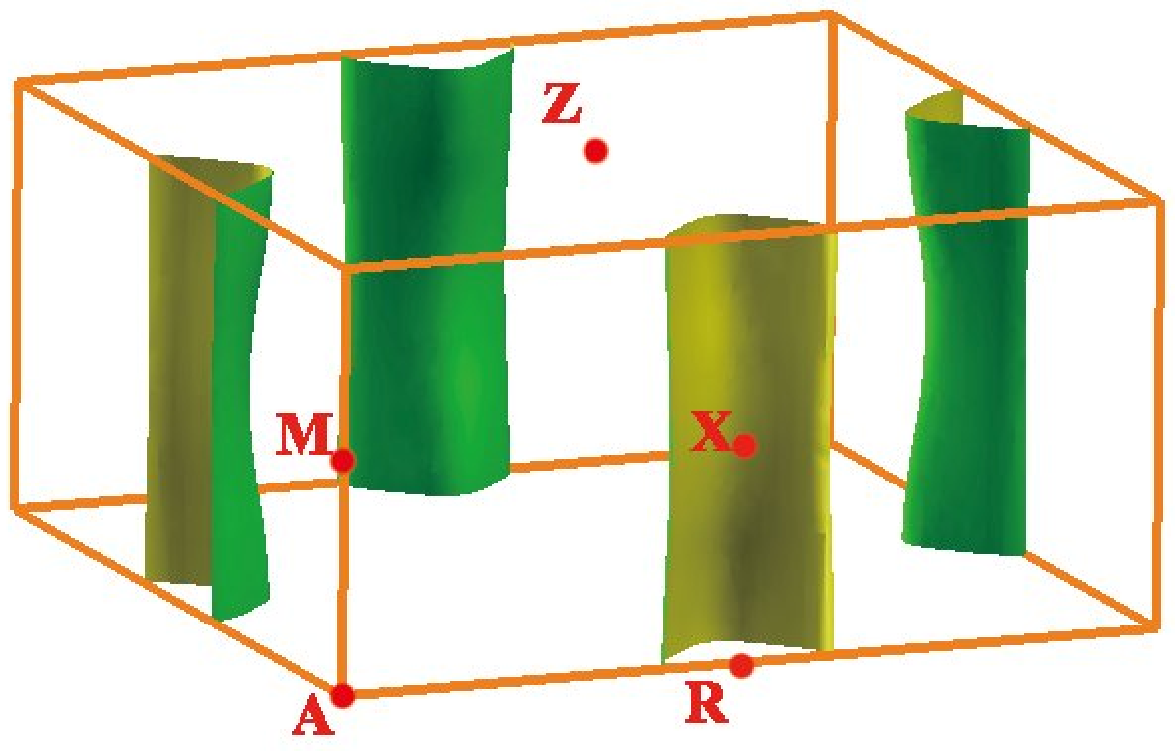}}%
 \subfigure[]{\label{fig:subfig:b} \includegraphics[width=0.450\textwidth]{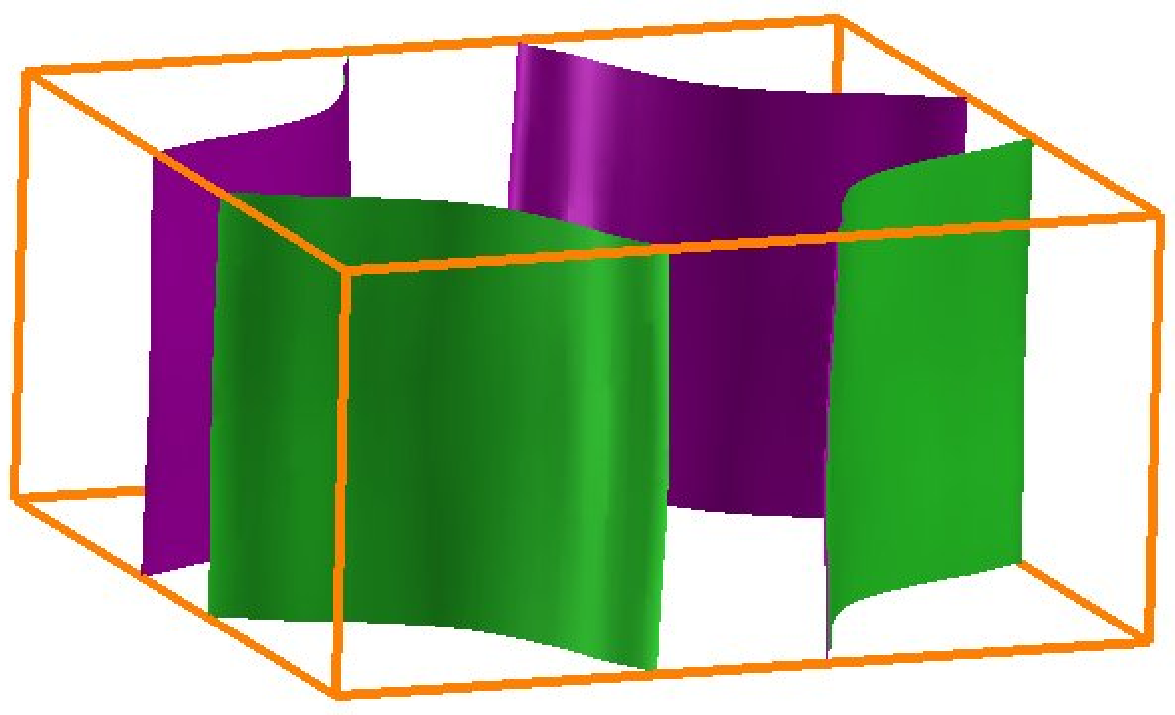}}\\
 \subfigure[]{\label{fig:subfig:c} \includegraphics[width=0.450\textwidth]{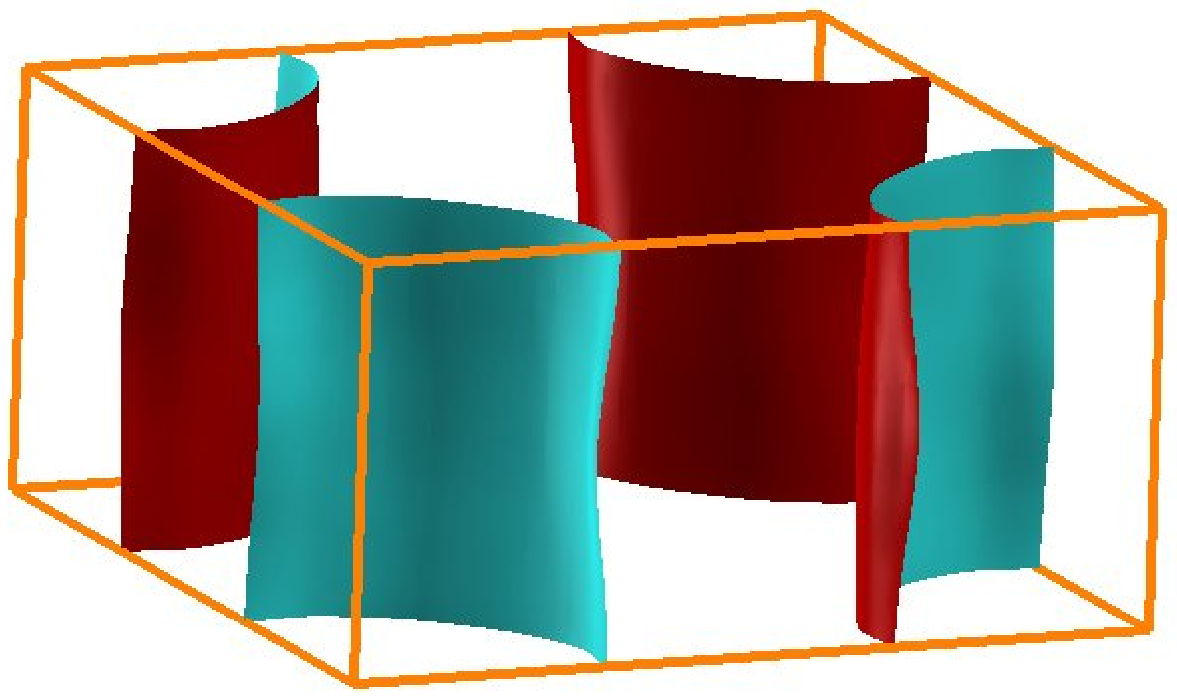}}%
 \subfigure[]{\label{fig:subfig:d} \includegraphics[width=0.450\textwidth]{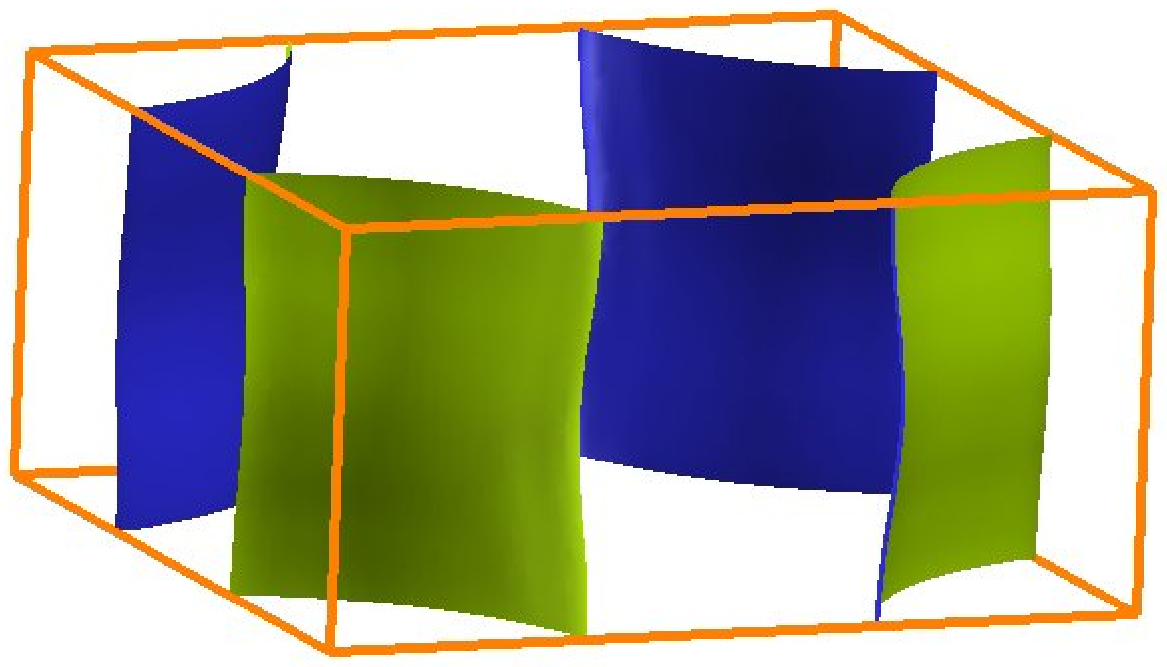}}%
\caption{\label{fig:Fs}(Color online) The Fermi surface sheet of
LaONiP shown in the first Brillouin zone centered around the
$\Gamma$ point.}
\end{figure*}

Now, Let's turn our attention to band structure. As shown in
Fig.~\ref{fig:band}, the characteristic feature of band structures
is the strongly pronounced two dimensionality along $\Gamma$-Z and
A-M. We observe four bands across the Fermi energy, which have been
indicated in the figure with different colors. The corresponding
Fermi surface in the first Brillouin zone are displayed in
Fig.~\ref{fig:Fs}. Due to the two-dimensional electronic structure,
all Fermi surfaces are cylindrical-like sheets parallel to the $k_z$
direction. The first sheet is hole cylinder centered along R-X
whereas the other three are electron cylinders centered along the
A-M high symmetry line. Such a series of Fermi surfaces are
different from the results of LaOFeP \cite{LaFeOPfermi} and
LaOFeAs\cite{AsFe1,corralted,ref1}, where five bands across the
Fermi level with four of them being cylindrical-like sheets(two
electron cylinders along A-M and the other two hole-like cylinders
along $\Gamma$-Z direction) and the other one being a
three-dimensional hole pocket along $\Gamma$-Z. The absence of
distorted sphere Fermi surface sheet suggests the two dimensionality
of band structure in LaONiP is more pronounced than the case of
LaOFePn. This can be explained by the fact that Ni$^{2+}$ (3d$^8$)
contribute two more electrons than Fe$^{2+}$ in LaOFePn , nickel
based system has relatively higher carrier density and the hole
bands tend to be filled. Thus, three-dimensional hole pocket
disappeared and electron bands dominate the conductivity. This is
also been evidenced by recent experiment \cite{LaONiAs}, in which
they found that the charge carriers of LaONiAs are dominantly
electron-type.

\begin{figure}
\centering
\includegraphics[width=0.380\textwidth]{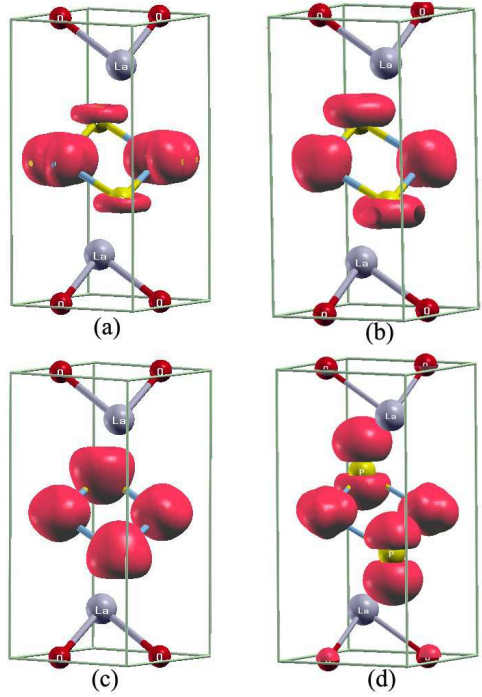}%
\caption{\label{fig:band_charg}(Color online) The band decomposed
charge density of four bands across the Fermi level. Iso surfaces
correspond to 0.035 $e/\AA^3$. In order to visualize the orbital
character of P atoms, we choose 0.025 $e/\AA^3$ for band \emph{a}.}
\end{figure}

\begin{figure}
\centering
\includegraphics[width=0.38\textwidth]{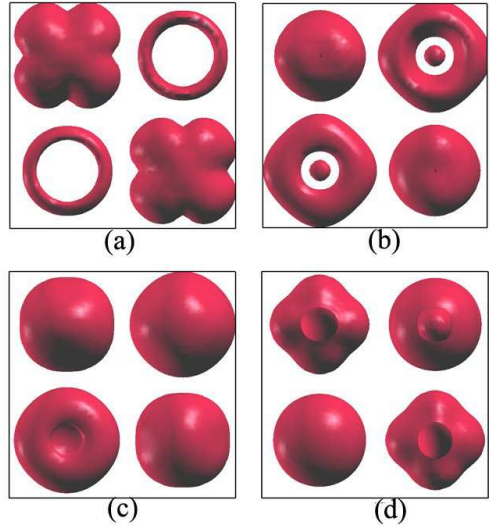}%
\caption{\label{fig:chg_TOP}(Color online) Top view of the
iso-surface of band decomposed charge density of four bands across
the Fermi level.Iso surfaces correspond to 0.035 $e/\AA^3$. In order
to visualize the orbital character of P atoms, we choose 0.025
$e/\AA^3$ for band \emph{a}.}
\end{figure}

For further insight into the mechanism, we give the band decomposed
charge density in Fig.~\ref{fig:band_charg}. Although the Fermi
surface of all bands have similar shape, the iso-surfaces of band
decomposed charge density are different from each other as shown in
Fig.~\ref{fig:band_charg} and Fig.~\ref{fig:chg_TOP}, which also
suggests that the orbital component is different. From the shape of
iso-surface, we can speculate that the band \emph{a} is made up of
Ni-$3d_{xy}$ and weakly mixture with $p-\sigma$ orbit,which may be
built with $p_x$ and $p_y$. In case of the band \emph{d}, the Ni
orbit is mainly contributed by $3d_{x^2-y^2}$ and the states of P
atom exhibit a $p_{z}$ orbital character. It seems that the orbital
component of band \emph{b} and \emph c is more complicated than
\emph a and \emph d. Combined with different orbital picture, we can
deduce the 3d orbital character of \emph c is mostly contribute by
$3d_{z^2}$ and maybe some component of $3d_{x^2-y^2}$. Whereas the
band \emph b may be mixed with $3d_{xz}$ and $3d_{yz}$, the states
of P are similar to \emph a,which may be composed of $p_x$ and
$p_y$. Our results shows that not only the $3d_{x^2-y^2}$ driving
superconductor in layered copper-based oxides, but also all five 3d
orbits are contribute to charge carriers at the Fermi level. So, it
seems that the underlying mechanism is very complicated in ZrCuSiAs
structured superconductor.

\section{\label{sec:conclusion} Conclusion}
Based on First-principles calculation, we have investigated the
electronic structure, binding properties and Fermi surface character
of a new ZrCuSiAs structured superconductor LaONiP.  Our results
indicate that the density of states and bonding properties are
similar to the results of LaOFeP. However, the Fermi surface
character exhibits remarkable difference and two-dimensional
character is more significant in case of LaONiP. We also find that
the orbital components of Fermi surfaces are very complicated, which
suggests that the superconductivity mechanism is remarkably
different from the well-known cuprate superconductors.
\begin{acknowledgments}
This work is supported in part by the Key Program of Educational
Department (No: 07A070) and Scientific Research Foundation of
Guangxi University (Grant No: X071117), China.
\end{acknowledgments}


\begin{thebibliography}{42}
\expandafter\ifx\csname
natexlab\endcsname\relax\def\natexlab#1{#1}\fi
\expandafter\ifx\csname bibnamefont\endcsname\relax
  \def\bibnamefont#1{#1}\fi
\expandafter\ifx\csname bibfnamefont\endcsname\relax
  \def\bibfnamefont#1{#1}\fi
\expandafter\ifx\csname citenamefont\endcsname\relax
  \def\citenamefont#1{#1}\fi
\expandafter\ifx\csname url\endcsname\relax
  \def\url#1{\texttt{#1}}\fi
\expandafter\ifx\csname urlprefix\endcsname\relax\def\urlprefix{URL
}\fi \providecommand{\bibinfo}[2]{#2}
\providecommand{\eprint}[2][]{\url{#2}}

\bibitem[{\citenamefont{Johnson and Jeitschko}(1974)}]{ZrCuSiAs}
\bibinfo{author}{\bibfnamefont{V.}~\bibnamefont{Johnson}} \bibnamefont{and}
  \bibinfo{author}{\bibfnamefont{W.}~\bibnamefont{Jeitschko}},
  \bibinfo{journal}{J. Solid State Chem.} \textbf{\bibinfo{volume}{11}},
  \bibinfo{pages}{161} (\bibinfo{year}{1974}).

\bibitem[{\citenamefont{Zimmer et~al.}(1995)\citenamefont{Zimmer, Jeitschko,
  Albering, Glaum, and Reehuis}}]{JAC-1995}
\bibinfo{author}{\bibfnamefont{B.~I.} \bibnamefont{Zimmer}},
  \bibinfo{author}{\bibfnamefont{W.}~\bibnamefont{Jeitschko}},
  \bibinfo{author}{\bibfnamefont{J.~H.} \bibnamefont{Albering}},
  \bibinfo{author}{\bibfnamefont{R.}~\bibnamefont{Glaum}}, \bibnamefont{and}
  \bibinfo{author}{\bibfnamefont{M.}~\bibnamefont{Reehuis}},
  \bibinfo{journal}{J. Alloy. Compd.} \textbf{\bibinfo{volume}{229}},
  \bibinfo{pages}{238} (\bibinfo{year}{1995}).

\bibitem[{\citenamefont{Kaczorowski et~al.}(1994)\citenamefont{Kaczorowski,
  Albering, Noel, and Jeitschko}}]{UcuPO}
\bibinfo{author}{\bibfnamefont{D.}~\bibnamefont{Kaczorowski}},
  \bibinfo{author}{\bibfnamefont{J.~H.} \bibnamefont{Albering}},
  \bibinfo{author}{\bibfnamefont{H.}~\bibnamefont{Noel}}, \bibnamefont{and}
  \bibinfo{author}{\bibfnamefont{W.}~\bibnamefont{Jeitschko}},
  \bibinfo{journal}{J. Alloy. Compd.} \textbf{\bibinfo{volume}{216}},
  \bibinfo{pages}{117} (\bibinfo{year}{1994}).

\bibitem[{\citenamefont{Albering and Jeitschko}(1996)}]{ThCuPO}
\bibinfo{author}{\bibfnamefont{J.~H.} \bibnamefont{Albering}} \bibnamefont{and}
  \bibinfo{author}{\bibfnamefont{W.}~\bibnamefont{Jeitschko}},
  \bibinfo{journal}{Z. Naturforsch., B: Chem. Sci.}
  \textbf{\bibinfo{volume}{51}}, \bibinfo{pages}{257} (\bibinfo{year}{1996}).

\bibitem[{\citenamefont{Nientiedt and Jeitschko}(1998)}]{LaZnOP}
\bibinfo{author}{\bibfnamefont{A.~T.} \bibnamefont{Nientiedt}}
  \bibnamefont{and}
  \bibinfo{author}{\bibfnamefont{W.}~\bibnamefont{Jeitschko}},
  \bibinfo{journal}{Inorg. Chem.} \textbf{\bibinfo{volume}{37}},
  \bibinfo{pages}{386} (\bibinfo{year}{1998}).

\bibitem[{\citenamefont{Takano et~al.}(2008)\citenamefont{Takano, Komatsuzaki,
  Komasaki, Watanabe, Takahashi, and Takase}}]{electronicLaZnOP}
\bibinfo{author}{\bibfnamefont{Y.}~\bibnamefont{Takano}},
  \bibinfo{author}{\bibfnamefont{S.}~\bibnamefont{Komatsuzaki}},
  \bibinfo{author}{\bibfnamefont{H.}~\bibnamefont{Komasaki}},
  \bibinfo{author}{\bibfnamefont{T.}~\bibnamefont{Watanabe}},
  \bibinfo{author}{\bibfnamefont{Y.}~\bibnamefont{Takahashi}},
  \bibnamefont{and} \bibinfo{author}{\bibfnamefont{K.}~\bibnamefont{Takase}},
  \bibinfo{journal}{J. Alloy. Compd.} \textbf{\bibinfo{volume}{451}},
  \bibinfo{pages}{467} (\bibinfo{year}{2008}).

\bibitem[{\citenamefont{Nientiedt et~al.}(1997)\citenamefont{Nientiedt,
  Jeitschko, Pollmeier, and Brylak}}]{LaMnPO1997qem}
\bibinfo{author}{\bibfnamefont{A.}~\bibnamefont{Nientiedt}},
  \bibinfo{author}{\bibfnamefont{W.}~\bibnamefont{Jeitschko}},
  \bibinfo{author}{\bibfnamefont{P.}~\bibnamefont{Pollmeier}},
  \bibnamefont{and} \bibinfo{author}{\bibfnamefont{M.}~\bibnamefont{Brylak}},
  \bibinfo{journal}{Z. Nat.Forsch., B J. Chem. Sci.}
  \textbf{\bibinfo{volume}{52}}, \bibinfo{pages}{560} (\bibinfo{year}{1997}).

\bibitem[{\citenamefont{Kabbour et~al.}(2005)\citenamefont{Kabbour, Cario, and
  Boucher}}]{kabbour2005rdn}
\bibinfo{author}{\bibfnamefont{H.}~\bibnamefont{Kabbour}},
  \bibinfo{author}{\bibfnamefont{L.}~\bibnamefont{Cario}}, \bibnamefont{and}
  \bibinfo{author}{\bibfnamefont{F.}~\bibnamefont{Boucher}},
  \bibinfo{journal}{J. Mater. Chem} \textbf{\bibinfo{volume}{15}},
  \bibinfo{pages}{3525} (\bibinfo{year}{2005}).

\bibitem[{\citenamefont{Kamihara et~al.}(2006)\citenamefont{Kamihara,
  Hiramatsu, Hirano, Kawamura, Yanagi, Kamiya, and Hosono}}]{jacs_LaFeOP}
\bibinfo{author}{\bibfnamefont{Y.}~\bibnamefont{Kamihara}},
  \bibinfo{author}{\bibfnamefont{H.}~\bibnamefont{Hiramatsu}},
  \bibinfo{author}{\bibfnamefont{M.}~\bibnamefont{Hirano}},
  \bibinfo{author}{\bibfnamefont{R.}~\bibnamefont{Kawamura}},
  \bibinfo{author}{\bibfnamefont{H.}~\bibnamefont{Yanagi}},
  \bibinfo{author}{\bibfnamefont{T.}~\bibnamefont{Kamiya}}, \bibnamefont{and}
  \bibinfo{author}{\bibfnamefont{H.}~\bibnamefont{Hosono}},
  \bibinfo{journal}{J. Am. Chem. Soc.} \textbf{\bibinfo{volume}{128}},
  \bibinfo{pages}{10012} (\bibinfo{year}{2006}).

\bibitem[{\citenamefont{Liang et~al.}(2007)\citenamefont{Liang, Che, Yang,
  Tian, Xiao, Lu, Li, and Li}}]{LaOFeP_SSt}
\bibinfo{author}{\bibfnamefont{C.~Y.} \bibnamefont{Liang}},
  \bibinfo{author}{\bibfnamefont{R.~C.} \bibnamefont{Che}},
  \bibinfo{author}{\bibfnamefont{H.~X.} \bibnamefont{Yang}},
  \bibinfo{author}{\bibfnamefont{H.~F.} \bibnamefont{Tian}},
  \bibinfo{author}{\bibfnamefont{R.~J.} \bibnamefont{Xiao}},
  \bibinfo{author}{\bibfnamefont{J.~B.} \bibnamefont{Lu}},
  \bibinfo{author}{\bibfnamefont{R.}~\bibnamefont{Li}}, \bibnamefont{and}
  \bibinfo{author}{\bibfnamefont{J.~Q.} \bibnamefont{Li}},
  \bibinfo{journal}{Supercond. Sci. Tech.} \textbf{\bibinfo{volume}{20}},
  \bibinfo{pages}{687} (\bibinfo{year}{2007}).

\bibitem[{\citenamefont{Kamihara et~al.}(2008)\citenamefont{Kamihara, Watanabe,
  Hirano, and Hosono}}]{LaOFeAs_F}
\bibinfo{author}{\bibfnamefont{Y.}~\bibnamefont{Kamihara}},
  \bibinfo{author}{\bibfnamefont{T.}~\bibnamefont{Watanabe}},
  \bibinfo{author}{\bibfnamefont{M.}~\bibnamefont{Hirano}}, \bibnamefont{and}
  \bibinfo{author}{\bibfnamefont{H.}~\bibnamefont{Hosono}},
  \bibinfo{journal}{J.Am. Chem. Soc.} \textbf{\bibinfo{volume}{130}},
  \bibinfo{pages}{3296} (\bibinfo{year}{2008}).

\bibitem[{\citenamefont{Wen et~al.}(2008)\citenamefont{Wen, Mu, Fang, Yang, and
  Zhu}}]{LaOFeAs_Sr}
\bibinfo{author}{\bibfnamefont{H.-H.} \bibnamefont{Wen}},
  \bibinfo{author}{\bibfnamefont{G.}~\bibnamefont{Mu}},
  \bibinfo{author}{\bibfnamefont{L.}~\bibnamefont{Fang}},
  \bibinfo{author}{\bibfnamefont{H.}~\bibnamefont{Yang}}, \bibnamefont{and}
  \bibinfo{author}{\bibfnamefont{X.}~\bibnamefont{Zhu}},
  \bibinfo{journal}{Europhys. Lett.} \textbf{\bibinfo{volume}{82}},
  \bibinfo{pages}{17009} (\bibinfo{year}{2008}).

\bibitem[{\citenamefont{Chen et~al.}(2008{\natexlab{a}})\citenamefont{Chen, Li,
  Wu, Li, Hu, Dong, Zheng, Luo, and Wang}}]{Ce-2008}
\bibinfo{author}{\bibfnamefont{G.~F.} \bibnamefont{Chen}},
  \bibinfo{author}{\bibfnamefont{Z.}~\bibnamefont{Li}},
  \bibinfo{author}{\bibfnamefont{D.}~\bibnamefont{Wu}},
  \bibinfo{author}{\bibfnamefont{G.}~\bibnamefont{Li}},
  \bibinfo{author}{\bibfnamefont{W.~Z.} \bibnamefont{Hu}},
  \bibinfo{author}{\bibfnamefont{J.}~\bibnamefont{Dong}},
  \bibinfo{author}{\bibfnamefont{P.}~\bibnamefont{Zheng}},
  \bibinfo{author}{\bibfnamefont{J.~L.} \bibnamefont{Luo}}, \bibnamefont{and}
  \bibinfo{author}{\bibfnamefont{N.~L.} \bibnamefont{Wang}},
  \bibinfo{journal}{arXiv:0803.3790}  (\bibinfo{year}{2008}{\natexlab{a}}).

\bibitem[{\citenamefont{Ren et~al.}(2008{\natexlab{a}})\citenamefont{Ren, Yang,
  Lu, Yi, Che, Dong, Sun, and Zhao}}]{Pr-2008}
\bibinfo{author}{\bibfnamefont{Z.-A.} \bibnamefont{Ren}},
  \bibinfo{author}{\bibfnamefont{J.}~\bibnamefont{Yang}},
  \bibinfo{author}{\bibfnamefont{W.}~\bibnamefont{Lu}},
  \bibinfo{author}{\bibfnamefont{W.}~\bibnamefont{Yi}},
  \bibinfo{author}{\bibfnamefont{G.-C.} \bibnamefont{Che}},
  \bibinfo{author}{\bibfnamefont{X.-L.} \bibnamefont{Dong}},
  \bibinfo{author}{\bibfnamefont{L.-L.} \bibnamefont{Sun}}, \bibnamefont{and}
  \bibinfo{author}{\bibfnamefont{Z.-X.} \bibnamefont{Zhao}},
  \bibinfo{journal}{arXiv:0803.4283}  (\bibinfo{year}{2008}{\natexlab{a}}).

\bibitem[{\citenamefont{Ren et~al.}(2008{\natexlab{b}})\citenamefont{Ren, Yang,
  Lu, Yi, Shen, Li, Che, Dong, Sun, Zhou et~al.}}]{Nd-2008}
\bibinfo{author}{\bibfnamefont{Z.-A.} \bibnamefont{Ren}},
  \bibinfo{author}{\bibfnamefont{J.}~\bibnamefont{Yang}},
  \bibinfo{author}{\bibfnamefont{W.}~\bibnamefont{Lu}},
  \bibinfo{author}{\bibfnamefont{W.}~\bibnamefont{Yi}},
  \bibinfo{author}{\bibfnamefont{X.-L.} \bibnamefont{Shen}},
  \bibinfo{author}{\bibfnamefont{Z.-C.} \bibnamefont{Li}},
  \bibinfo{author}{\bibfnamefont{G.-C.} \bibnamefont{Che}},
  \bibinfo{author}{\bibfnamefont{X.-L.} \bibnamefont{Dong}},
  \bibinfo{author}{\bibfnamefont{L.-L.} \bibnamefont{Sun}},
  \bibinfo{author}{\bibfnamefont{F.}~\bibnamefont{Zhou}}, \bibnamefont{et~al.},
  \bibinfo{journal}{arXiv:0803.4234}  (\bibinfo{year}{2008}{\natexlab{b}}).

\bibitem[{\citenamefont{Chen et~al.}(2008{\natexlab{b}})\citenamefont{Chen, Wu,
  Wu, Liu, Chen, and Fang}}]{sm-2008}
\bibinfo{author}{\bibfnamefont{X.~H.} \bibnamefont{Chen}},
  \bibinfo{author}{\bibfnamefont{T.}~\bibnamefont{Wu}},
  \bibinfo{author}{\bibfnamefont{G.}~\bibnamefont{Wu}},
  \bibinfo{author}{\bibfnamefont{R.~H.} \bibnamefont{Liu}},
  \bibinfo{author}{\bibfnamefont{H.}~\bibnamefont{Chen}}, \bibnamefont{and}
  \bibinfo{author}{\bibfnamefont{D.~F.} \bibnamefont{Fang}},
  \bibinfo{journal}{arXiv:0803.3603}  (\bibinfo{year}{2008}{\natexlab{b}}).

\bibitem[{\citenamefont{Ren et~al.}(2008{\natexlab{c}})\citenamefont{Ren, Che,
  Dong, Yang, Lu, Yi, Shen, Li, Sun, Zhou et~al.}}]{sm1-2008}
\bibinfo{author}{\bibfnamefont{Z.-A.} \bibnamefont{Ren}},
  \bibinfo{author}{\bibfnamefont{G.-C.} \bibnamefont{Che}},
  \bibinfo{author}{\bibfnamefont{X.-L.} \bibnamefont{Dong}},
  \bibinfo{author}{\bibfnamefont{J.}~\bibnamefont{Yang}},
  \bibinfo{author}{\bibfnamefont{W.}~\bibnamefont{Lu}},
  \bibinfo{author}{\bibfnamefont{W.}~\bibnamefont{Yi}},
  \bibinfo{author}{\bibfnamefont{X.-L.} \bibnamefont{Shen}},
  \bibinfo{author}{\bibfnamefont{Z.-C.} \bibnamefont{Li}},
  \bibinfo{author}{\bibfnamefont{L.-L.} \bibnamefont{Sun}},
  \bibinfo{author}{\bibfnamefont{F.}~\bibnamefont{Zhou}}, \bibnamefont{et~al.},
  \bibinfo{journal}{arXiv:0804.2582}  (\bibinfo{year}{2008}{\natexlab{c}}).

\bibitem[{\citenamefont{Cheng et~al.}(2008)\citenamefont{Cheng, Fang, Yang,
  Zhu, Mu, Luo, Wang, and Wen}}]{Gd-2008-51}
\bibinfo{author}{\bibfnamefont{P.}~\bibnamefont{Cheng}},
  \bibinfo{author}{\bibfnamefont{L.}~\bibnamefont{Fang}},
  \bibinfo{author}{\bibfnamefont{H.}~\bibnamefont{Yang}},
  \bibinfo{author}{\bibfnamefont{X.}~\bibnamefont{Zhu}},
  \bibinfo{author}{\bibfnamefont{G.}~\bibnamefont{Mu}},
  \bibinfo{author}{\bibfnamefont{H.}~\bibnamefont{Luo}},
  \bibinfo{author}{\bibfnamefont{Z.}~\bibnamefont{Wang}}, \bibnamefont{and}
  \bibinfo{author}{\bibfnamefont{H.-H.} \bibnamefont{Wen}},
  \bibinfo{journal}{SCIENCE IN CHINA G} \textbf{\bibinfo{volume}{51}},
  \bibinfo{pages}{719} (\bibinfo{year}{2008}).

\bibitem[{\citenamefont{Krellner et~al.}(2007)\citenamefont{Krellner, Kini,
  Bruning, Koch, Rosner, Nicklas, Baenitz, and Geibel}}]{CeRuPO}
\bibinfo{author}{\bibfnamefont{C.}~\bibnamefont{Krellner}},
  \bibinfo{author}{\bibfnamefont{N.~S.} \bibnamefont{Kini}},
  \bibinfo{author}{\bibfnamefont{E.~M.} \bibnamefont{Bruning}},
  \bibinfo{author}{\bibfnamefont{K.}~\bibnamefont{Koch}},
  \bibinfo{author}{\bibfnamefont{H.}~\bibnamefont{Rosner}},
  \bibinfo{author}{\bibfnamefont{M.}~\bibnamefont{Nicklas}},
  \bibinfo{author}{\bibfnamefont{M.}~\bibnamefont{Baenitz}}, \bibnamefont{and}
  \bibinfo{author}{\bibfnamefont{C.}~\bibnamefont{Geibel}},
  \bibinfo{journal}{Phys. Rev. B} \textbf{\bibinfo{volume}{76}},
  \bibinfo{pages}{104418} (\bibinfo{year}{2007}).

\bibitem[{\citenamefont{Krellner and Geibel}(2007)}]{krellner2007scg}
\bibinfo{author}{\bibfnamefont{C.}~\bibnamefont{Krellner}} \bibnamefont{and}
  \bibinfo{author}{\bibfnamefont{C.}~\bibnamefont{Geibel}},
  \bibinfo{journal}{arXiv: 0709.4144}  (\bibinfo{year}{2007}).

\bibitem[{\citenamefont{Watanabe et~al.}(2007)\citenamefont{Watanabe, Yanagi,
  Kamiya, Kamihara, Hiramatsu, Hirano, and Hosono}}]{LaNiO_IC}
\bibinfo{author}{\bibfnamefont{T.}~\bibnamefont{Watanabe}},
  \bibinfo{author}{\bibfnamefont{H.}~\bibnamefont{Yanagi}},
  \bibinfo{author}{\bibfnamefont{T.}~\bibnamefont{Kamiya}},
  \bibinfo{author}{\bibfnamefont{Y.}~\bibnamefont{Kamihara}},
  \bibinfo{author}{\bibfnamefont{H.}~\bibnamefont{Hiramatsu}},
  \bibinfo{author}{\bibfnamefont{M.}~\bibnamefont{Hirano}}, \bibnamefont{and}
  \bibinfo{author}{\bibfnamefont{H.}~\bibnamefont{Hosono}},
  \bibinfo{journal}{Inorg. Chem.} \textbf{\bibinfo{volume}{46}},
  \bibinfo{pages}{7719} (\bibinfo{year}{2007}).

\bibitem[{\citenamefont{Tegel et~al.}(2008)\citenamefont{Tegel, Bichler, and
  Johrendt}}]{LaNiPO_SSC}
\bibinfo{author}{\bibfnamefont{M.}~\bibnamefont{Tegel}},
  \bibinfo{author}{\bibfnamefont{D.}~\bibnamefont{Bichler}}, \bibnamefont{and}
  \bibinfo{author}{\bibfnamefont{D.}~\bibnamefont{Johrendt}},
  \bibinfo{journal}{Solid State Sci.} \textbf{\bibinfo{volume}{10}},
  \bibinfo{pages}{193} (\bibinfo{year}{2008}).

\bibitem[{\citenamefont{Li et~al.}(2008)\citenamefont{Li, Chen, Dong, Li, Hu,
  Wu, Su, Zheng, Xiang, Wang et~al.}}]{LaONiAs}
\bibinfo{author}{\bibfnamefont{Z.}~\bibnamefont{Li}},
  \bibinfo{author}{\bibfnamefont{G.~F.} \bibnamefont{Chen}},
  \bibinfo{author}{\bibfnamefont{J.}~\bibnamefont{Dong}},
  \bibinfo{author}{\bibfnamefont{G.}~\bibnamefont{Li}},
  \bibinfo{author}{\bibfnamefont{W.~Z.} \bibnamefont{Hu}},
  \bibinfo{author}{\bibfnamefont{D.}~\bibnamefont{Wu}},
  \bibinfo{author}{\bibfnamefont{S.~K.} \bibnamefont{Su}},
  \bibinfo{author}{\bibfnamefont{P.}~\bibnamefont{Zheng}},
  \bibinfo{author}{\bibfnamefont{T.}~\bibnamefont{Xiang}},
  \bibinfo{author}{\bibfnamefont{N.~L.} \bibnamefont{Wang}},
  \bibnamefont{et~al.}, \bibinfo{journal}{arXiv:0803.2572}
  (\bibinfo{year}{2008}).

\bibitem[{\citenamefont{Kresse and
  Furthm\"uller}(1996{\natexlab{a}})}]{vasp_CMS}
\bibinfo{author}{\bibfnamefont{G.}~\bibnamefont{Kresse}} \bibnamefont{and}
  \bibinfo{author}{\bibfnamefont{J.}~\bibnamefont{Furthm\"uller}},
  \bibinfo{journal}{Comp. Mater. Sci.} \textbf{\bibinfo{volume}{6}},
  \bibinfo{pages}{15} (\bibinfo{year}{1996}{\natexlab{a}}).

\bibitem[{\citenamefont{Kresse and
  Furthm\"uller}(1996{\natexlab{b}})}]{vaspPhysRevB.54.11169}
\bibinfo{author}{\bibfnamefont{G.}~\bibnamefont{Kresse}} \bibnamefont{and}
  \bibinfo{author}{\bibfnamefont{J.}~\bibnamefont{Furthm\"uller}},
  \bibinfo{journal}{Phys. Rev. B} \textbf{\bibinfo{volume}{54}},
  \bibinfo{pages}{11169} (\bibinfo{year}{1996}{\natexlab{b}}).

\bibitem[{\citenamefont{Bl\"ochl}(1994)}]{vasp_PAW}
\bibinfo{author}{\bibfnamefont{P.~E.} \bibnamefont{Bl\"ochl}},
  \bibinfo{journal}{Phys. Rev. B} \textbf{\bibinfo{volume}{50}},
  \bibinfo{pages}{17953} (\bibinfo{year}{1994}).

\bibitem[{\citenamefont{Kresse and Joubert}(1999)}]{vasppawPhysRevB.59.1758}
\bibinfo{author}{\bibfnamefont{G.}~\bibnamefont{Kresse}} \bibnamefont{and}
  \bibinfo{author}{\bibfnamefont{D.}~\bibnamefont{Joubert}},
  \bibinfo{journal}{Phys. Rev. B} \textbf{\bibinfo{volume}{59}},
  \bibinfo{pages}{1758} (\bibinfo{year}{1999}).

\bibitem[{\citenamefont{Perdew and Wang}(1992)}]{PW91}
\bibinfo{author}{\bibfnamefont{J.~P.} \bibnamefont{Perdew}} \bibnamefont{and}
  \bibinfo{author}{\bibfnamefont{Y.}~\bibnamefont{Wang}},
  \bibinfo{journal}{Phys. Rev. B} \textbf{\bibinfo{volume}{45}},
  \bibinfo{pages}{13244} (\bibinfo{year}{1992}).

\bibitem[{\citenamefont{Monkhorst and Pack}(1976)}]{VASP_MP}
\bibinfo{author}{\bibfnamefont{H.~J.} \bibnamefont{Monkhorst}}
  \bibnamefont{and} \bibinfo{author}{\bibfnamefont{J.~D.} \bibnamefont{Pack}},
  \bibinfo{journal}{Phys. Rev. B} \textbf{\bibinfo{volume}{13}},
  \bibinfo{pages}{5188} (\bibinfo{year}{1976}).

\bibitem[{\citenamefont{Bl\"ochl et~al.}(1994)\citenamefont{Bl\"ochl, Jepsen,
  and Andersen}}]{PhysRevB.49.16223}
\bibinfo{author}{\bibfnamefont{P.~E.} \bibnamefont{Bl\"ochl}},
  \bibinfo{author}{\bibfnamefont{O.}~\bibnamefont{Jepsen}}, \bibnamefont{and}
  \bibinfo{author}{\bibfnamefont{O.~K.} \bibnamefont{Andersen}},
  \bibinfo{journal}{Phys. Rev. B} \textbf{\bibinfo{volume}{49}},
  \bibinfo{pages}{16223} (\bibinfo{year}{1994}).

\bibitem[{\citenamefont{Kokalj}(2003)}]{xcrysden}
\bibinfo{author}{\bibfnamefont{A.}~\bibnamefont{Kokalj}},
  \bibinfo{journal}{Comp. Mater. Sci.} \textbf{\bibinfo{volume}{28}},
  \bibinfo{pages}{155} (\bibinfo{year}{2003}), \bibinfo{note}{code available
  from http://www.xcrysden.org/.}

\bibitem[{\citenamefont{Boeri et~al.}(2008)\citenamefont{Boeri, Dolgov, and
  Golubov}}]{ref1}
\bibinfo{author}{\bibfnamefont{L.}~\bibnamefont{Boeri}},
  \bibinfo{author}{\bibfnamefont{O.~V.} \bibnamefont{Dolgov}},
  \bibnamefont{and} \bibinfo{author}{\bibfnamefont{A.~A.}
  \bibnamefont{Golubov}}, \bibinfo{journal}{arXiv:0803.2703}
  (\bibinfo{year}{2008}).

\bibitem[{\citenamefont{Singh and Du}(2008)}]{AsFe1}
\bibinfo{author}{\bibfnamefont{D.~J.} \bibnamefont{Singh}} \bibnamefont{and}
  \bibinfo{author}{\bibfnamefont{M.~H.} \bibnamefont{Du}},
  \bibinfo{journal}{arXiv:0803.0429}  (\bibinfo{year}{2008}).

\bibitem[{\citenamefont{Yin et~al.}(2008)\citenamefont{Yin, Lebegue, Han, Neal,
  Savrasov, and Pickett}}]{ref2}
\bibinfo{author}{\bibfnamefont{Z.~P.} \bibnamefont{Yin}},
  \bibinfo{author}{\bibfnamefont{S.}~\bibnamefont{Lebegue}},
  \bibinfo{author}{\bibfnamefont{M.~J.} \bibnamefont{Han}},
  \bibinfo{author}{\bibfnamefont{B.}~\bibnamefont{Neal}},
  \bibinfo{author}{\bibfnamefont{S.~Y.} \bibnamefont{Savrasov}},
  \bibnamefont{and} \bibinfo{author}{\bibfnamefont{W.~E.}
  \bibnamefont{Pickett}}, \bibinfo{journal}{arXiv:0804.3355}
  (\bibinfo{year}{2008}).

\bibitem[{\citenamefont{Lebegue}(2007)}]{LaFeOPfermi}
\bibinfo{author}{\bibfnamefont{S.}~\bibnamefont{Lebegue}},
  \bibinfo{journal}{Phys. Rev. B} \textbf{\bibinfo{volume}{75}},
  \bibinfo{pages}{035110} (\bibinfo{year}{2007}).

\bibitem[{\citenamefont{Haule et~al.}(2008)\citenamefont{Haule, Shim, and
  Kotliar}}]{corralted}
\bibinfo{author}{\bibfnamefont{K.}~\bibnamefont{Haule}},
  \bibinfo{author}{\bibfnamefont{J.~H.} \bibnamefont{Shim}}, \bibnamefont{and}
  \bibinfo{author}{\bibfnamefont{G.}~\bibnamefont{Kotliar}},
  \bibinfo{journal}{arXiv:0803.1279}  (\bibinfo{year}{2008}).

\bibitem[{\citenamefont{Yaghi et~al.}(2003)\citenamefont{Yaghi, O'Keeffe,
  Ockwig, Chae, Eddaoudi, and Kim}}]{10.1038/nature01650}
\bibinfo{author}{\bibfnamefont{O.~M.} \bibnamefont{Yaghi}},
  \bibinfo{author}{\bibfnamefont{M.}~\bibnamefont{O'Keeffe}},
  \bibinfo{author}{\bibfnamefont{N.~W.} \bibnamefont{Ockwig}},
  \bibinfo{author}{\bibfnamefont{H.~K.} \bibnamefont{Chae}},
  \bibinfo{author}{\bibfnamefont{M.}~\bibnamefont{Eddaoudi}}, \bibnamefont{and}
  \bibinfo{author}{\bibfnamefont{J.}~\bibnamefont{Kim}},
  \bibinfo{journal}{Nature} \textbf{\bibinfo{volume}{423}},
  \bibinfo{pages}{705} (\bibinfo{year}{2003}).

\bibitem[{\citenamefont{Chen et~al.}(2001)\citenamefont{Chen, Eddaoudi, Hyde,
  O'Keeffe, and Yaghi}}]{BanglinChen02092001}
\bibinfo{author}{\bibfnamefont{B.}~\bibnamefont{Chen}},
  \bibinfo{author}{\bibfnamefont{M.}~\bibnamefont{Eddaoudi}},
  \bibinfo{author}{\bibfnamefont{S.~T.} \bibnamefont{Hyde}},
  \bibinfo{author}{\bibfnamefont{M.}~\bibnamefont{O'Keeffe}}, \bibnamefont{and}
  \bibinfo{author}{\bibfnamefont{O.~M.} \bibnamefont{Yaghi}},
  \bibinfo{journal}{Science} \textbf{\bibinfo{volume}{291}},
  \bibinfo{pages}{1021} (\bibinfo{year}{2001}).

\bibitem[{\citenamefont{Cario et~al.}(2005)\citenamefont{Cario, Kabbour, and
  Meerschaut}}]{CarioL.cm048180p}
\bibinfo{author}{\bibfnamefont{L.}~\bibnamefont{Cario}},
  \bibinfo{author}{\bibfnamefont{H.}~\bibnamefont{Kabbour}}, \bibnamefont{and}
  \bibinfo{author}{\bibfnamefont{A.}~\bibnamefont{Meerschaut}},
  \bibinfo{journal}{Chem. Mater.} \textbf{\bibinfo{volume}{17}},
  \bibinfo{pages}{234} (\bibinfo{year}{2005}).

\bibitem[{\citenamefont{Kabbour et~al.}(2006)\citenamefont{Kabbour, Cario,
  Danot, and Meerschaut}}]{KabbourH.ic051592v}
\bibinfo{author}{\bibfnamefont{H.}~\bibnamefont{Kabbour}},
  \bibinfo{author}{\bibfnamefont{L.}~\bibnamefont{Cario}},
  \bibinfo{author}{\bibfnamefont{M.}~\bibnamefont{Danot}}, \bibnamefont{and}
  \bibinfo{author}{\bibfnamefont{A.}~\bibnamefont{Meerschaut}},
  \bibinfo{journal}{Inorg. Chem.} \textbf{\bibinfo{volume}{45}},
  \bibinfo{pages}{917} (\bibinfo{year}{2006}).

\bibitem[{\citenamefont{Bader}(1990)}]{bader1990amq}
\bibinfo{author}{\bibfnamefont{R.~F.~W.} \bibnamefont{Bader}},
  \emph{\bibinfo{title}{{Atoms in Molecules: A Quantum Theory}}}
  (\bibinfo{publisher}{Clarendon Press}, \bibinfo{year}{1990}).

\bibitem[{\citenamefont{Henkelman et~al.}(2006)\citenamefont{Henkelman,
  Arnaldsson, and J{\'o}nsson}}]{bader2006far}
\bibinfo{author}{\bibfnamefont{G.}~\bibnamefont{Henkelman}},
  \bibinfo{author}{\bibfnamefont{A.}~\bibnamefont{Arnaldsson}},
  \bibnamefont{and}
  \bibinfo{author}{\bibfnamefont{H.}~\bibnamefont{J{\'o}nsson}},
  \bibinfo{journal}{Comp. Mater. Sci.} \textbf{\bibinfo{volume}{36}},
  \bibinfo{pages}{354} (\bibinfo{year}{2006}).

\end{thebibliography}

%
\end{document}